
\headline={\ifnum\pageno=1\firstheadline\else
\ifodd\pageno\rightheadline \else\leftheadline\fi\fi}
\def\firstheadline{February, \ 1992 \hss ROM2F-93/5}
\def\rightheadline{\hfil}
\def\leftheadline{\hfil}
\pageno=1
\footline={\ifnum\pageno=1\firstfootline\else\otherfootline\fi}
\def\firstfootline{}
\def\otherfootline{\rm\hss -- \folio \ -- \hss}
\font\tenbf=cmbx10
\font\tenrm=cmr10
\font\tenit=cmti10
\font\elevenbf=cmbx10 scaled\magstep 1
\font\elevenrm=cmr10 scaled\magstep 1
\font\elevenit=cmti10 scaled\magstep 1

\line{\hfil }
\vglue 1cm
\hsize=6.0truein
\vsize=8.5truein
\parindent=3pc
\baselineskip=10pt
\centerline{\tenbf ANOMALY \ CANCELLATIONS \ AND \ OPEN - STRING \ THEORIES}
\vglue 1.0cm
\centerline{\tenrm Augusto \ \ SAGNOTTI }
\baselineskip=13pt
\centerline{\tenit Dipartimento di Fisica, Universit\`a di Roma ``Tor
Vergata''}
 \baselineskip=12pt
\centerline{\tenit I.N.F.N, Sezione di Roma ``Tor
Vergata''}
\baselineskip=12pt
\centerline{\tenit Via della Ricerca Scientfica, 1 \ \ \ 00133  \ Roma \ \
ITALY} \vglue 0.3cm
\vglue 1.5cm
\centerline{\tenrm ABSTRACT}
\vglue 0.3cm
{\rightskip=3pc
 \leftskip=3pc
 \tenrm\baselineskip=12pt
 \noindent
In models of oriented closed strings, anomaly cancellations are deeply linked
to the {\it modular invariance} of the torus amplitude.  If open
and/or unoriented strings are allowed, there are no non-trivial modular
transformations in the additional genus-one amplitudes (Klein bottle,
annulus and M\"obius strip).  As originally recognized by Green and Schwarz,
in the ten-dimensional type-I superstring the anomaly
cancellation results from a delicate interplay between the
contributions of these additional surfaces.  In lower-dimensional models, the
possible presence of a number of antisymmetric tensors yields a
generalization of the Green-Schwarz mechanism.  I illustrate these results by
referring to some six-dimensional chiral models, and I conclude by addressing
the additional difficulties that one meets when trying to extend the
construction to chiral four-dimensional models.
\vglue 0.8cm }
\line{\elevenbf 1. Introduction \hfil}
\bigskip
\smallskip
\baselineskip=14pt
\elevenrm
In ten dimensions, the unique Lagrangian of $N=1$ supergravity coupled to
$N=1$ supersymmetric Yang-Mills theory [1] describes the low-energy
interactions of the massless modes of two vastly different string models, the
heterotic string [2] and the type-I superstring.  As a result,
the anomaly cancellations at work in the two models
take a unique form, the Green-Schwarz mechanism [3].  There are two crucial
ingredients in this mechanism. The first, a proper choice of
gauge group, does not suffice to eliminate the whole anomaly polynomial.
Rather, it disposes {\elevenit only} of the terms containing irreducible traces
of six gauge-field strengths and of six
Riemann tensors.  The second, key ingredient, is the presence
of an antisymmetric two-tensor, that is to acquire proper transformations
under gauge and Lorentz symmetries. One may then
exhibit a local counterterm that would suffice to cancel the residual
anomaly polynomial, in this case the product of an eight form and a four form.

Though quite suggestive, the low-energy analysis should be supplemented by a
proper study of string amplitudes.  The crucial differences between the
two string models are then evident. In the heterotic string, a model of
oriented
closed strings only, the {\elevenit modular invariance} of the torus amplitude
may be held responsible for the cancellation of the residual
anomaly.  A neat discussion of this
result may be found in ref. [4].  On the other hand, in the type-I superstring,
containing both unoriented closed strings and open strings, three additional
surfaces (the Klein bottle, the annulus and the M\"obius strip) contribute
to the anomaly.  The cancellation results in this case from a
delicate interplay
between the contributions of these three surfaces that, by a
suitable choice of ``time'' coordinate on the world sheet, may
{\elevenit all}
be associated to the propagation of closed strings [3,5,6]. A related, crucial
feature of the
type-I superstring has to do with the Chan-Paton construction of the
gauge symmetry [7,8]. This is the origin of
long-known restrictions on the gauge group [8]: all exceptional groups
are excluded, even $E_8 \otimes
E_8$, despite its being allowed by the low-energy analysis. The lesson
that should be drawn from all this is that lower-dimensional
models are likely to exhibit further crucial differences between the
two classes of string theories.

The study of open-string models has long been hampered by the lack of a
procedure to construct new interesting solutions. In ref. [9] I proposed
that consistent open-string models should be defined in terms of an
orbifold-like construction in parameter space, and I argued that
they should be somehow associated to arbitrary closed-string models with
a symmetry under the interchange of left and right modes.
In a series of subsequent
papers [10,11,12] these observations were turned into an algorithm to construct
rational open-string theories.  Cardy's analysis of the annulus amplitude
in rational conformal field theory [13] proved
to be the key to achieve Chan-Paton symmetry breaking.

A procedure that uses closed-string models as the starting point
has the potential to lead to some surprising results.
Thus, the chiral six-dimensional
models of ref. [11] contain in their spectra a number of (anti)self-dual
two-tensors that draw their origin from Ramond-Ramond sectors of the
``parent'' type-IIb string.  I have no good argument to justify the presence
of these fields starting from the type-I theory, since the key feature of
the construction {\elevenit is} using the closed-string theory as the
starting point.  Still, the antisymmetric tensors
play a crucial role in the anomaly cancellation procedure [14], and give it
some distinctive features that I would like to discuss in the remainder
of this talk.  Other, related, developments are reviewed in ref. [15].

\bigskip
\bigskip
\line{\elevenbf 2. Six - Dimensional Chiral Models and the Generalized
Mechanism \hfil}
\bigskip
\smallskip
I will begin by discussing two classes of six-dimensional models.  They are
obtained starting from different rational models describing the
compactification of the type-IIb superstring on  $K_3$.
In the first class of models [11], the torus amplitude is
$$
T \ = \ \sum_{i=1}^8 \ | \chi_i |^2 \ + \ \sum_{i=1}^8 \ | \tilde{\chi}_i |^2
\qquad ,
\eqno(2.1)
$$
where the sixteen generalized characters are
suitable combinations of the four characters of $SO(4)$ level one.  Defining
$Q_o \ = \ V O - C C$, $Q_v \ = \ O V - S S$, $Q_s \ = \ O C - S O$ and
$Q_c \ = \ V S - C V$, the sixteen generalized characters are
$$
\eqalignno{
&\chi_{1} \ = \ Q_{O} OO \; \ + \ Q_{V} V V \quad , \qquad
\tilde{\chi}_{1} \ = \ Q_{S} SO \; \ + \ Q_{C} CV \quad ,\cr
&\chi_{2} \ = \ Q_{O} OV \; \ + \ Q_{V} VO \quad , \qquad
\tilde{\chi}_{2} \ = \ Q_{S} SV \; \ + \ Q_{C} CO \quad ,\cr
&\chi_{3} \ = \ Q_{O} CC \; \ + \ Q_{V} SS \quad , \qquad
\tilde{\chi}_{3} \ = \ Q_{S} VC \; \ + \ Q_{C} OS \quad ,\cr
&\chi_{4} \ = \ Q_{O} CS \; \ + \ Q_{V} SC \quad , \qquad
\tilde{\chi}_{4} \ = \ Q_{S} VS \; \ + \ Q_{C} OC \quad ,\cr
&\chi_{5} \ = \ Q_{O} VV \; \ + \ Q_{V} OO \quad , \qquad
\tilde{\chi}_{5} \ = \ Q_{S} CV \; \ + \ Q_{C} SO \quad ,\cr
&\chi_{6} \ = \ Q_{O} VO \; \ + \ Q_{V} OV \quad , \qquad
\tilde{\chi}_{6} \ = \ Q_{S} CO \; \ + \ Q_{C} SV \quad ,\cr
&\chi_{7} \ = \ Q_{O} SS \; \ + \ Q_{V} CC \quad , \qquad
\tilde{\chi}_{7} \ = \ Q_{S} OS \; \ + \ Q_{C} VC \quad ,\cr
&\chi_{8} \ = \ Q_{O} SC \; \ + \ Q_{V} CS \quad , \qquad
\tilde{\chi}_{8} \ = \ Q_{S} OC \; \ + \ Q_{C} VS
\quad . \qquad &(2.2)}
$$
The model has $N=2$ supersymmetry in six dimensions (thus, it would have
$N=4$ supersymmetry if trivially reduced to four dimensions), and all
massless modes are associated to the terms $| \chi_1 |^2$,
$| \chi_5 |^2$, $| \tilde{\chi}_1 |^2$, $| \tilde{\chi}_6 |^2$,
$| \tilde{\chi}_7 |^2$ and $| \tilde{\chi}_8 |^2$.  The massless spectrum
contains a supergravity multiplet (in the notation of ref. [16]
the $N=4b$ multiplet) and a total of 21 tensor multiplets.  This field
content is fixed completely by the anomaly analysis, as pointed out
in refs. [17,18].  Starting from this model and proceeding as in ref. [11],
one may derive a class of open-string ``descendants'' whose anomaly polynomials
do not contain any irreducible traces, provided the Chan-Paton multiplicities
satisfy the tadpole conditions
$$
\eqalignno{
\sum_{i=1}^8 \ n_i \ &= \ \sum_{i=1}^8 \ \tilde{n}_i \ = \ 16 \qquad ; \cr
n_5 \ - n_1 \ + \ \tilde{n}_1 \ &+ \ \tilde{n}_6 \ + \ \tilde{n}_7 \ + \
\tilde{n}_8 \  = \ 8 \qquad ; \cr
n_6 \ - n_2 \ + \ \tilde{n}_2 \ &+ \ \tilde{n}_5 \ + \ \tilde{n}_7 \ + \
\tilde{n}_8 \  = \ 8 \qquad ; \cr
n_7 \ - n_3 \ + \ \tilde{n}_3 \ &+ \ \tilde{n}_5 \ + \ \tilde{n}_6 \ + \
\tilde{n}_8 \  = \ 8 \qquad ; \cr
n_8 \ - n_4 \ + \ \tilde{n}_4 \ &+ \ \tilde{n}_5 \ + \ \tilde{n}_6 \ + \
\tilde{n}_7 \  = \ 8 \qquad . &(2.3) \cr}
$$
These allow, for instance, a $USp(8)^4$ gauge group.  The resulting massless
spectrum contains
chiral fermions in the representations ({\bf 8},{\bf 1},{\bf 1},{\bf 8}),
({\bf 1},{\bf 8},{\bf 8},{\bf 1}), ({\bf 8},{\bf 1},{\bf 8},{\bf 1}) and
({\bf 1},{\bf 8},{\bf 1},{\bf 8}).  In addition to the scalar
multiplets containing these fermions, the massless spectrum contains
the $N=2b$ supergravity multiplet, five tensor multiplets and sixteen scalar
multiplets from the closed sector, as well as the gauge multiplet from the
open sector.  I would like to stress that the model contains a number
of antisymmetric two-tensors, to wit five self-dual tensors from the tensor
multiplets and one antiself-dual tensor from the $N=2b$ supergravity
multiplet.  This is rather fortunate since, even after imposing the
tadpole
conditions of eq. (3), the anomaly polynomial {\elevenit does not}
factorize.  Therefore, in these models the Green-Schwarz mechanism
may not work in the standard
fashion.  For instance, for the $USp(8)^4$ model one finds the
residual polynomial
$$
\eqalignno{\rm{A} \ = \ {1 \over 8} \ &\Big\lbrace \ ( \rm{tr} {F_1}^2
)^2 \ + \ ( \rm{tr} {F_2}^2 )^2 \ + \ ( \rm{tr} {F_{\tilde{7}}}^2 )^2 \ + \ (
\rm{tr} {F_{\tilde{8}}}^2 )^2 \ \Big\rbrace \cr
+ \ {1 \over 16} \ &\Big\lbrace \ \rm{tr} {F_1}^2 \ + \
\rm{tr} {F_2}^2 \ + \ \rm{tr} {F_{\tilde{7}}}^2 \ + \ \rm{tr} {F_{\tilde{8}}}^2
\Big\rbrace \ \rm{tr} R^2 \cr
- {1 \over 4} &\Big\lbrace \ \rm{tr} {F_1}^2 \ \rm{tr} {F_{\tilde{7}}}^2 \ + \
\rm{tr} {F_1}^2 \ \rm{tr} {F_{\tilde{8}}}^2 \ + \ \rm{tr} {F_2}^2 \ \rm{tr}
{F_{\tilde{7}}}^2 \ + \
\rm{tr} {F_2}^2 \ \rm{tr} {F_{\tilde{8}}}^2 \ \Big\rbrace \cr
&- \ {1 \over 32} \ ( \rm{tr} R^2 )^2 \ \qquad \qquad , &(2.4) \cr }
$$
where the two-forms are defined as follows:
$R^{ab} = {1 \over 2} {R_{\mu \nu}}^{ab} dx^{\mu} dx^{\nu}$ and
$F^{a} = {1 \over 2} {F_{\mu \nu}}^{a} dx^{\mu} dx^{\nu}$.
If the polynomial is diagonalized, the end result is rather pleasing, since
there are precisely {\elevenit six} non-zero eigenvalues, as many
as the antisymmetric tensors, and
$$
\eqalignno{\rm{A} \ = \ &- \ {1 \over 32} \ {\Big\lbrace \
\rm{tr} {F_1}^2 \ + \ \rm{tr} {F_2}^2 \ + \  \rm{tr} {F_{\tilde{7}}}^2 \ + \
\rm{tr} {F_{\tilde{8}}}^2 \ - \ \rm{tr} R^2 \ \Big\rbrace}^2 \cr
&+ \ {3 \over 32} \ {\Big\lbrace \
\rm{tr} {F_1}^2 \ + \ \rm{tr} {F_2}^2 \ - \  \rm{tr} {F_{\tilde{7}}}^2 \ - \
\rm{tr} {F_{\tilde{8}}}^2 \  \Big\rbrace}^2 \cr
&+ \ {1 \over 32} \ {\Big\lbrace \
\rm{tr} {F_1}^2 \ - \ \rm{tr} {F_2}^2 \ + \
\rm{tr} {F_{\tilde{7}}}^2 \ - \ \rm{tr} {F_{\tilde{8}}}^2 \
 \Big\rbrace}^2 \cr
&+ \ {1 \over 32} \ {\Big\lbrace \
\rm{tr} {F_1}^2 \ - \ \rm{tr} {F_2}^2 \ - \
\rm{tr} {F_{\tilde{7}}}^2 \ + \ \rm{tr} {F_{\tilde{8}}}^2 \
 \Big\rbrace}^2 \qquad . &(2.5) \cr
 }
$$

It should be appreciated that only one combination
contains the gravitational two-form.
Strictly speaking, only the different normalization
of the second term suggests that the quadratic form has six non-zero
eigenvalues.  This, however, may be seen quite clearly
if all sixteen charge sectors are allowed, subject only to the
tadpole conditions of eq. (3).
The various combinations of field traces
correspond precisely to the rows of the $S$ matrix
acting on the sixteen characters of eq. (2) that identify the sectors
containing the antisymmetric tensors, and
$$
{\rm A} \ = \ - \ {1 \over 2} {\big\{ \sum_m \ S_{1m} \ trF_m^2 \
- \ 4 \ trR^2 \big\}}^2
\ + \ {1 \over 2} \sum_k \ {\big\{ \sum_m \ S_{km} \ trF_m^2 \big\}}^2
\qquad ,
\eqno(2.6)
$$
where $k=5,\tilde{1},\tilde{6},\tilde{7},\tilde{8}$.
The first line contains the
sum of all gauge-field strenghts and is the only one containing the
Riemann curvature. In addition, the various contributions
enter the anomaly polynomial in a way that corresponds to a
Minkowski metric with signature $(1-n)$. If, following standard
practice, the
eight-form in eq. (6) is converted into a Green-Schwarz counterterm
$$
\Delta L \ = \ + \ {1 \over 2} \sum\limits_{ij} \ \eta_{ij} \ F^{(i)} B^{(j)}
\qquad ,
\eqno(2.7)
$$
where $F^{(i)}$ denote combinations of Yang-Mills (and gravitational)
curvatures,
the modified field strengths for the antisymmetric tensors are
$$
H^{(i)} \ = \ d B^{(i)} \ + \ \omega^{(i)} \qquad ,
\eqno(2.8)
$$
with $\omega^{(i)}$ proper combinations of Yang-Mills and
gravitational Chern-Simons forms. The couplings between combinations of
Yang-Mills Chern-Simons forms and antisymmetric tensors
may be explored by constructing the field equations of the
low-energy theory (there are long-standing problems with action principles
for (anti)self-dual bosons [18]).  On the other hand,
as in ten dimensions, the coupling
to the gravitational Chern-Simons form is not in the
low-energy field theory.  The counterterm of eq. (7)
has precisely the $SO(1,n)$ symmetry that one would expect
to be present in this class of supergravity models [16].

Before displaying the structure of the generalized Chern-Simons couplings,
I would like to
repeat the exercise for another class of models discussed
in ref. [11].  In this case the torus amplitude is
$$
\eqalignno{T \ = \ |\chi_1|^2 \ &+ \ |\chi_2|^2 \ + \ |\chi_5|^2 \ + \
|\chi_6|^2 \ + \ {\rm ``tilde''} \cr
{\chi_3} \bar{\chi}_4 \ &+ \ \chi_4 \bar{\chi}_3 \ + \
\chi_7 \bar{\chi}_8 \ + \ \chi_8 \bar{\chi}_7 \ +
\ {\rm ``tilde''} \qquad , \qquad &(2.9) \cr}
$$
where ``tilde'' stands for the corresponding characters from the ``twisted''
sector.
Following the procedure described in ref. [11] one may construct a
class of open-string ``descendants'' whose annulus amplitudes are
built out eight composite characters, $\chi_1 + \chi_2$,
$\chi_3 + \chi_4$, $\chi_5 + \chi_6$, $\chi_7 + \chi_8$, and
the corresponding ones from the ``twisted'' sector.  The resulting
anomaly polynomials do not
contain irreducible traces, provided the eight independent Chan-Paton
multiplicities satisfy the tadpole conditions
$$\eqalignno{
\sum_i \ n_i \ = \ \sum_i \ \tilde{n}_i \ &= \ 8 \cr
n_1 \ + \ n_2 \ - \ n_3 \ - \ n_4 \ = \
\tilde{n}_3 \ &+ \ \tilde{n}_4 \ - \ \tilde{n}_1 \ - \ \tilde{n}_2 \cr
n_1 \ - \ n_2 \ - \ n_3 \ + \ n_4 \ = \
\tilde{n}_3 \ &- \ \tilde{n}_4 \ - \ \tilde{n}_1 \ + \ \tilde{n}_2
\qquad . \qquad &(2.10) \cr}
$$
For instance, one may choose a gauge group $USp(4)^4$.
Then, apart from the gaugini, the resulting model contains
chiral fermions in the representations
({\bf 4},{\bf 1},{\bf 4},{\bf 1}) and
({\bf 1},{\bf 4},{\bf 1},{\bf 4}),
as well as two families in each of the representations
({\bf 4},{\bf 1},{\bf 1},{\bf 4}) and
({\bf 1},{\bf 4},{\bf 4},{\bf 1}). In addition to the scalar
multiplets containing these fermions, the massless spectrum contains
the $N=2b$ supergravity multiplet, seven tensor multiplets and fourteen scalar
multiplets from the closed sector, as well as the gauge multiplet from the
open sector.
In this case there are only eight types of quantum numbers, and the residual
anomaly polynomial may be written
$$
\eqalignno{\rm{A} \ = \ &- \ {1 \over 16} \ {\Big\lbrace \
\sum_i \ \rm{tr} {F_i}^2 \ + \ \sum_i \ \rm{tr} {F_{\tilde{i}}}^2 \ - \
{1 \over 2} \ \rm{tr} R^2 \ \Big\rbrace}^2 \cr
&+ \ {1 \over 16} \ {\Big\lbrace \
\sum_i \ \rm{tr} {F_i}^2 \ - \ \sum_i \ \rm{tr} {\tilde{F}_i}^2 \
\Big\rbrace}^2 \cr
&+ \ {1 \over 16} \ {\Big\lbrace \
\rm{tr} {F_1}^2 \ + \ \rm{tr} {F_2}^2 \ - \
\rm{tr} {F_{3}}^2 \ - \ \rm{tr} {F_{4}}^2 \ + \ {\rm ``tilde''} \
 \Big\rbrace}^2 \cr
&+ \ {1 \over 16} \ {\Big\lbrace \
\rm{tr} {F_1}^2 \ - \ \rm{tr} {F_2}^2 \ - \
\rm{tr} {F_{3}}^2 \ + \ \rm{tr} {F_{4}}^2 \ - \ {\rm ``tilde''} \
 \Big\rbrace}^2 \qquad . &(2.11) \cr
 }
$$

Why is the Green-Schwarz mechanism using only four of the
eight available two-tensors?  This may be understood quite naturally
in terms of the construction of ref. [11].  Indeed, {\elevenit only} eight
of the sixteen characters are allowed in the transverse annulus amplitude.
They are $\chi_1$, $\chi_2$, $\chi_5$, $\chi_6$, and the
corresponding characters
from the twisted sector.  Out of these, only $\chi_1$, $\chi_5$,
$\tilde{\chi}_1$ and $\tilde{\chi}_6$ yield antisymmetric two-tensors.
Thus, the model uses precisely the four antisymmetric tensors that are
allowed in the vacuum channel of the annulus. These are the only ones
that may take part in the Green-Schwarz mechanism.

It is amusing to construct the field equations for $N=2b$ supergravity
coupled to $n$ tensor multiplets and to gauge multiplets via the
generalized Chern-Simons couplings required by the
anomaly analysis.  These equations extend the previous work of ref. [16], where
the authors considered the two cases of $n$ tensor multiplets with no
gauge multiplets and of a single tensor multiplet.\footnote*
{In this case the self-dual
two-tensor joins the antiself-dual one in the $N=2b$ supergravity
multiplet to yield a single two-tensor with no self-duality, and one may
write a Lagrangian in standard form.}
The low-energy supergravities corresponding to the two classes
of open-string models I have described contain matter multiplets as
well, but these coupled equations suffice
to display the structure of the generalized couplings.

In the notation of ref. [14], the $n$ scalar fields parametrize
the coset space $SO(1,n)/SO(n)$, and are conveniently described
using the $SO(1,n)$ matrix
$$
V \ = \ \pmatrix{v_0&v_M\cr {x^m}_0&{x^m}_M \cr}		\qquad .
\eqno(2.12)
$$
Out of the elements of $V$ one may construct the composite $SO(n)$ connection
$$
{S_{\mu}}^{[mn]} \ = \ ( \ \partial_{\mu} \ {x^m}_r \ ) \ {{\tilde{x}}^r}_{\ n}
\qquad ,
\eqno(2.13)
$$
antisymmetric in $(m,n)$ because of the (pseudo)orthogonal nature
of $V$.  The scalar kinetic term is then built out of
$$
{P^m}_{\mu} \ = \ \sqrt{{1 \over 2}} \ ( \ \partial_{\mu} \ v_r \ ) \
{{\tilde{x}}^r}_{\ m}	\qquad ,
\eqno(2.14)
$$
where $P$ satisfies $D_{[{\mu}}  {P^m}_{{\nu}]} \ = \ 0$.
The model contains $(n+1)$ tensor fields ${A^r}_{\mu \nu}$
that transform in the fundamental representation of $SO(1,n)$.  Combining
these fields and the Chern-Simons forms one may construct
the field strengths
$$
F^r \ = \ d A^r \ - \ c^{rz} \ \omega_z
\eqno(2.15)
$$
and, from these,
$$
\eqalign{ H_{\mu \nu \rho} \ &= \ v_r \ {F^r}_{\mu \nu \rho} \qquad , \cr
{K^m}_{\mu \nu \rho} \ &= \ {x^m}_r \ {F^r}_{\mu \nu \rho}\qquad . \cr}
\eqno(2.16)
$$
The spinor fields are a left-handed gravitino $\psi_{\mu}$, $n$ right-handed
spinors $\chi^m$ from the tensor multiplets and the gaugini
$\lambda$.  All spinors are $Sp(2)$ Majorana-Weyl.

The field equations of the spinor fields are
$$
\eqalign{ &{\gamma}^{\mu \nu \rho}  D_{\nu} \psi_{\rho} \ + \
H^{\mu \nu \rho} \ \gamma_{\nu} \psi_{\rho} \ - \ { i \over 2} \
K^{m \mu \nu \rho} \ \gamma_{\nu \rho} \chi^m \ - \
{i \over \sqrt{2}} \ {P^m}_{\nu} \ \gamma^{\nu} \gamma^{\mu}
 \chi^m  \cr
&- \  {1 \over {2 \sqrt{2}}} \ \gamma^{\sigma \tau} \ {\gamma^{\mu}} \
v_r \ c^{rz} \ {\rm tr}_z ( F^{\sigma \tau} \ \lambda ) \ = \ 0 \qquad , \cr
&\gamma^{\mu} D_{\mu} \chi^m \ - \ {1 \over 12} \ H_{\mu \nu \rho} \
\gamma^{\mu \nu \rho} \chi^m \ - \ {i \over 2} \ K^{m \mu \nu \rho} \
\gamma_{\mu \nu}  \psi_{\rho} \ + \ {i \over \sqrt{2}} \
{P^m}_{\nu} \ {\gamma}^{\mu}  {\gamma}^{\nu} \psi_{\mu} \cr
&- \ {i \over {2 \sqrt{2}}} \ {x^m}_r \ c^{rz} \ {\rm tr}_z ( \gamma^{\mu \nu}
\lambda
\ F_{\mu \nu} )
\ = \ 0  \qquad , \cr
&( v_r c^{rz}) \ \gamma^{\mu} D_{\mu} \lambda \ + \ {1 \over {\sqrt{2}}}
{P^m}_{\mu} ({x^m}_r c^{rz}) \ \gamma^{\mu} \lambda \ + \
{1 \over {2 \sqrt{2}}} \ ( v_r c^{rz}) \ F_{\lambda \tau} \gamma^{\mu}
\gamma^{\lambda
\tau} \psi_{\mu} \cr &+ \ {i \over {2 \sqrt{2}}} \ ({x^m}_r c^{rz}) \
\gamma^{\mu \nu}
\chi^m \ F_{\mu \nu} \ = \ 0  \qquad , \cr}
\eqno(2.17)
$$
while the field equations of the Bose fields are
$$
\eqalignno{
&D_{\mu}  P^{m \mu} \ - \ {\sqrt{2} \over 3} \  H^{\mu \nu \rho}
 {K^m}_{\mu \nu \rho} \ + \ {1 \over {2 \sqrt{2}}} \ {x^m}_r c^{rz} \
{\rm tr}_z ( F_{\alpha \beta} F^{\alpha \beta} ) \ = \ 0 \qquad . \cr
&R_{\mu \nu} \ - \ {1 \over 2} g_{\mu \nu}  R \ - \  H_{\mu \rho \sigma}
{H_{\nu}}^{\rho \sigma} \ - \ {K^m}_{\mu \rho \sigma}
{{K^m}_{\nu}}^{\rho \sigma} \ - \ 2 {P^m}_{\mu} {P^m}_{\nu} \cr &+  \
g_{\mu \nu} \ {P^m}_{\rho} P^{m \rho} \ + \
2 v_r \ c^{rz} \ {\rm tr}_z ( F_{\lambda \mu} {F^{\lambda}}_{\nu} \ - \
{1 \over 4} g_{\mu \nu} F^2 ) \ = \ 0  \qquad , \cr
&( v_r c^{rz}) \ D^{\mu} F_{\mu \nu} \ + \ \sqrt{2} \ ({x^m}_r c^{rz}) \
P^{m \mu} F_{\mu \nu}
- \ ( v_r c^{rz}) \ F_{\rho \sigma} {H_{\nu}}^{\rho \sigma} \cr &- \
({x^m}_r c^{rz}) \ F^{\rho \sigma} {K^m}_{\nu \rho \sigma} \ = \ 0
\qquad , &(2.18)\cr}
$$
together with the (anti)self-duality conditions for the antisymmetric
two-tensors, that read
$$
\eqalignno{ H_{\mu \nu \rho} \ &= \ {\tilde{H}}_{\mu \nu \rho} \cr
{K^m}_{\mu \nu \rho} \ &= \ - \ {{\tilde{K}}^m}_{\ \ \mu \nu \rho}
\qquad . &(2.19) \cr }
$$
A good consistency check comes from the supersymmetry transformations,
$$
\eqalignno{ &\delta  {e_{\mu}}^m \ = \ - \ i \ \bar{\epsilon}  \gamma^m
 \psi_{\mu} \qquad , \cr
&\delta \psi_{\mu} \ = \ D_{\mu} \ \epsilon \ + \ {1 \over 4}
\ H_{\mu \nu \rho} \ \gamma^{\nu \rho}  \epsilon \qquad , \cr
&\delta {A^r}_{\mu \nu} \ = \ i \ {\tilde{v}}^r \ {\bar{\psi}}_{[ \mu}
\gamma_{ \nu ]}  \epsilon \ - \ {1 \over 2} \ {{\tilde{x}}^r}_m \
{\bar{\chi}}^m  \gamma_{\mu \nu} \epsilon
- \ c^{rz} \ {\rm {tr}_z} ( A_{[ \mu} \ \delta
	A_{\nu ]} ) \qquad , \cr
&\delta \chi^m \ = \ - \ {i \over \sqrt{2}} \ {\gamma}^{\mu} {P^m}_{\mu} \
\epsilon \ + \ {i \over 12} \ {K_{\mu \nu \rho}}^m \ {\gamma}^{\mu \nu \rho}
\epsilon \qquad , \cr
&\delta v_r \ = \ {x^m}_r \ \bar{\epsilon} {\chi}^m
\qquad , \cr
&\delta \lambda \ = \ - \ {1 \over {2 \sqrt{2}}} \ F_{\mu \nu} \
{\gamma}^{\mu \nu} \epsilon \qquad , \cr
&\delta A_{\mu} \ = \ - \ {i \over {\sqrt{2}}} ( \bar{\epsilon} \gamma_{\mu}
\lambda )
\qquad \qquad . &(2.20) \cr}
$$
that close on the Bose fields in terms of all local symmetries in the model.
Under the transformations of eq. (20) the fermionic field equations turn
into the bosonic ones, as discussed in ref. [14]. The constants $c^{rz}$
determine the generalized Chern-Simons couplings and the effective gauge
charges of the vectors, thus limiting the effective range of the scalar
fields.
As in ref. [20], all these equations have been constructed to lowest order
in the spinor fields.  The methods of ref. [21] should provide
a convenient way of completing the construction.
\bigskip
\bigskip
\line{\elevenbf 3. Four-dimensional Models \hfil}
\bigskip
\medskip
I would like to conclude by pointing out the type of
difficulties one meets when trying to construct chiral four-dimensional
models.  A convenient class of four-dimensional models may be used to
illustrate the nature
of the problem.  The ``parent'' closed strings are obtained in this case as
$Z_2 \otimes Z_2$ orbifolds
of the type-IIb superstring compactified to four dimensions on the $SO(12)$
torus.  I will confine my attention to two choices for the torus partition
function,
$$
T_1 \ = \ \sum_i \ | \chi_i |^2 \qquad ,
\eqno(3.1)
$$
the diagonal modular invariant, and
$$
T_2 \ = \ \sum_{i,j} \ C_{ij} \ \chi_i \ {\bar{\chi}}_j \qquad ,
\eqno(3.2)
$$
the ``charge-conjugation'' modular invariant.  The ${\chi}_i$ are a
set of 64 generalized characters, and the label $i$ is a shorthand for
a triple of indices.  The first index, taking the values $o,v,s,c$,
relates the characters to projections and/or twistings of the four
$SO(12)$ characters $O_{12}, V_{12}, S_{12}, C_{12}$.  The last
two indices, taking the values $o,g,h,f$, are the usual labels for the
four sectors and for the four projections.  Thus, for instance,
$$
\eqalignno{\chi_{o,oo} \ &= \ \tau_{oo} OOO \ +  \ \tau_{og} VVO
 \ + \ \tau_{oh} OVV  \ + \ \tau_{of}  VOV  \qquad , \cr
\chi_{v,oo} \  &= \ \tau_{oo}  OVO  \ + \  \tau_{og}  VOO
 \ + \ \tau_{oh} OOV \ + \ \tau_{of} VVV \qquad ,
&(3.3) \cr}
$$
while
$$
\eqalignno{
\chi_{o,go}  \ &= \ \tau_{go}  OOS  \ + \ \tau_{gg}  VVS
 \ + \ \tau_{gh}  OVC \  + \ \tau_{gf}  VOC  \qquad , \cr
\chi_{v,go} \ &= \ \tau_{go}  OVS  \ + \  \tau_{gg}  VOS
 \ + \  \tau_{gh}  OOC \ + \ \tau_{gf}  VVC \qquad . &(3.4) \cr}
$$
In eqs. (3.3) and (3.4) the space-time characters are
$$
\eqalignno{\tau_{oo} \ &= \ vooo + ovvv - scsc - cscs \qquad
\tau_{og} \ = \ ovoo + vovv - sccs - cssc \cr
\tau_{oh} \ &= \ ooov + vvvo - ccss - sscc \qquad
\tau_{of} \ = \ oovo + vvov - ssss - cccc \cr
\tau_{go} \ &= \ voss + ovcc - scvo - csov \qquad
\tau_{gg} \ = \ ovss + vocc - scov - csvo \cr
\tau_{gh} \ &= \ oosc + vvcs - ccvv - ssoo \qquad
\tau_{gf} \ = \ oocs + vvsc - ssvv - ccoo \cr
\tau_{ho} \ &= \ vsso + occv - sovc - cvos \qquad
\tau_{hg} \ = \ ocso + vscv - soos - cvvc \cr
\tau_{hh} \ &= \ ossv + vcco - covs - svoc \qquad
\tau_{hf} \ = \ osco + vcsv - svvs - cooc \cr
\tau_{fo} \ &= \ vsvs + ococ - soco - cvsv \qquad
\tau_{fg} \ = \ ocvs + vsoc - sosv - cvco \cr
\tau_{fh} \ &= \ osvc + vcos - cocv - svso \qquad
\tau_{ff} \ = \ osos + vcvc - svcv - coso
\qquad . &(3.5) \cr}
$$
I will confine my attention to one choice of ``open-string Wilson lines''
on the M\"obius strip [11] such that the $P$ matrix acts as follows:
$$
\eqalignno{
\chi_{o,oo} &\rightarrow {1 \over 2} \ \big(
\ - \ \chi_{o,oo} \ + \ \chi_{o,og} \ + \ \chi_{o,oh} \ + \ \chi_{o,of} \
\big) \cr
\chi_{o,og} &\rightarrow {1 \over 2} \ \big( \
\ \chi_{o,oo} \ - \ \chi_{o,og} \ + \ \chi_{o,oh} \ + \ \chi_{o,of} \
\big) \cr
\chi_{o,oh} &\rightarrow {1 \over 2} \ \big( \
 \ \chi_{o,oo} \ + \ \chi_{o,og} \ - \ \chi_{o,oh} \ + \ \chi_{o,of} \
\big) \cr
\chi_{o,of} &\rightarrow {1 \over 2} \ \big( \
 \ \chi_{o,oo} \ + \ \chi_{o,og} \ + \ \chi_{o,oh} \ - \ \chi_{o,of} \
\big) \qquad . &(3.6) \cr}
$$
Other consistent choices of the $P$ matrix lead to similar conclusions.

If the torus partition function is the one given in eq. (1) the vacuum
channel of the annulus may only accommodate the $16$ ``untwisted'' characters.
As a result, the identity of the annulus is associated to the sum
of the four characters $\chi_{o,oo}$, $\chi_{o,og}$, $\chi_{o,oh}$
and $\chi_{o,of}$.  Given the first of eqs. (6), this is quite consistent
with the structure of the $P$ matrix, since in this case
the transverse M\"obius channel may only accommodate the identity.  There
is no problem with the tadpole conditions, and the
resulting open-string models have $N=1$ supersymmetry in four dimensions
but {\elevenit are not} chiral, on account of the structure of the identity.
On the other hand, if the torus invariant is the one given in eq. (2), all
$64$ characters may flow in the annulus amplitude.  The models are chiral,
have a large Chan-Paton charge space,
and one may show that the tadpole conditions from the ``twisted'' sectors,
easy to solve, imply the cancellation of all gauge
anomalies.\footnote*{These tadpole conditions are homogeneous, since
all chiral fermions are in fundamental representations of unitary groups.}
The trouble comes from the four tadpole conditions from the ``untwisted''
sector.  In this case all four characters in eq. (6) are allowed in the
vacuum channels of the Klein bottle and M\"obius amplitudes, and one finds
the conditions
$$
\eqalignno{&n_o \ + \ n_g \ + \ n_h \ + \ n_f \ = \ \ 32 \cr
&n_o \ + \ n_g \ - \ n_h \ - \ n_f \ = \ - 32 \cr
&n_o \ - \ n_g \ + \ n_h \ - \ n_f \ = \ - 32 \cr
&n_o \ - \ n_g \ - \ n_h \ + \ n_f \ = \ - 32  \qquad , &(3.3) \cr}
$$
where $n_i$ denote the total Chan-Paton multiplicities from the four
sectors of the spectrum.  The different
numbers of ``minus'' signs in the two sides of eq. (3),
necessary in order to
obtain a consistent symmetrization of the annulus amplitude,
force some of the multiplicities to be
negative, and therefore one may not impose these tadpole conditions in a
consitent fashion.  Though not directly related to four-dimensional
anomalies, from a two-dimensional viewpoint these conditions have a
dignity comparable to the others coming from the ``twisted'' sectors,
since they are also related to the decoupling of spurious states [6] from
vacuum channels.  For the time being, one is therefore forced to live
with other, less handy cases, where the available gauge groups of chiral
models with $N=1$ supersymmetry are small, typically products of $U(2)$
factors.  I will stop here, leaving a proper discussion of these models
to a future work.
\vglue 0.4cm
\vglue 0.6cm
\line{\elevenbf 4. Acknowledgements \hfil}
\vglue 0.4cm
I would like to thank the Organizers for their kind invitation,
while apologizing for being unable to attend the Meeting.  I am also
grateful to Massimo Bianchi and Gianfranco Pradisi for several discussions
on four-dimensional open-string models. This work was supported in
part by E.C. contract SCI-0394-C.
\vfill
\eject
\line{\elevenbf 5. References \hfil}
\bigskip
\item{1.} E. Bergshoeff, M. de Roo, B. de Wit and P. van
Nieuwenhuizen,  \hfill \break {\elevenit Nucl. Phys.} {\elevenbf B195} (1982)
97;\hfill \break  G.F. Chapline and N.S. Manton, {\elevenit Phys. Lett.}
{\elevenbf 120B} (1983) 105.
\item{2.}D.J. Gross, J.A. Harvey, E. Martinec and R. Rohm, \hfill \break
{\elevenit Nucl. Phys.} {\elevenbf B256} (1985) 253.
\item{3.} M.B. Green and J.H. Schwarz, {\elevenit Phys. Lett.}
{\elevenbf 149B} (1984) 117.
\item{4.} A.N. Schellekens and N.P. Warner, {\elevenit Nucl. Phys.}
{\elevenbf B287} (1987) 317.
\item{5.} M.B. Green and J.H. Schwarz, {\elevenit Phys. Lett.}
{\elevenbf 151B} (1985) 21.
\item{6.} J. Polchinski and Y. Cai, {\elevenit Nucl. Phys.}
{\elevenbf B296} (1988) 91.
\item{7.} H.M. Chan and J.E. Paton, {\elevenit Nucl. Phys.}
{\elevenbf B10} (1969) 516.
\item{8.}J.H. Schwarz, {\elevenit in} ``Current Problems in Particle
Theory'', \hfill \break
Proc. J. Hopkins Conf. {\elevenbf 6} (Florence, 1982);
\hfill \break
N. Marcus and A. Sagnotti, {\elevenit Phys. Lett.}
{\elevenbf 119B} (1982) 97; {\elevenbf 188B} (1987) 58.
\item{9.} A. Sagnotti, {\elevenit in} Cargese '87, ``Non-Perturbative
Quantum Field Theory'', \hfill \break
eds. G. Mack {\elevenit et al} (Plenum Press, 1988), p. 521.
\item{10.} M. Bianchi and A. Sagnotti, {\elevenit Phys. Lett.}
{\elevenbf 211B} (1988) 407; \hfill \break {\elevenit Phys. Lett.}
{\elevenbf 237B} (1989) 389; \hfill \break
G. Pradisi and A. Sagnotti, {\elevenit Phys. Lett.}
{\elevenbf 216B} (1989) 59.
\item{11.} M. Bianchi and A. Sagnotti, {\elevenit Phys. Lett.}
{\elevenbf 247B} (1990) 517; \hfill \break
{\elevenit Nucl. Phys.}{\elevenbf B361} (1991) 519.
\item{12.} M. Bianchi, G. Pradisi and A. Sagnotti, {\elevenit Phys. Lett.}
{\elevenbf 273B} (1991) 389; \hfill \break
{\elevenit Nucl. Phys.}{\elevenbf B376} (1992) 365.
\item{13.} J. Cardy, {\elevenit Nucl. Phys.}{\elevenbf B324} (1989) 581.
\item{14.} A. Sagnotti, {\elevenit Phys. Lett.}
{\elevenbf 294B} (1992) 196.
\item{15.} G. Pradisi and A. Sagnotti, {\elevenbf ROM2F}-92/53 ,
Proc. Tenth Annual Meeting on General Relativity (World Scientific, to appear).
\item{16.} L.J. Romans, {\elevenit Nucl. Phys.} {\elevenbf B276}
(1986) 71; \hfill \break
H. Nishino and E. Sezgin, {\elevenit Nucl. Phys.} {\elevenbf B278}
(1986) 353.
\item{17.} N. Seiberg, {\elevenit Nucl. Phys.} {\elevenbf B303}
(1988) 286.
\item{18.} L. Alvarez-Gaum\'e and E. Witten, {\elevenit Nucl. Phys.}
{\elevenbf B34} (1983) 269.
\item{19.} N. Marcus and J.H. Schwarz, {\elevenit Phys. Lett.}
{\elevenbf 115B} (1982) 111.
\item{20.} J.H. Schwarz, {\elevenit Nucl. Phys.}
{\elevenbf B226} (1983) 289.
\item{21.} P.S. Howe and P.C. West, {\elevenit Nucl. Phys.}
{\elevenbf B238} (1984) 181.
\eject
\bye